\documentclass[journal,onecolumn]{IEEEtran}

\ifCLASSINFOpdf
\else
\fi

\usepackage{graphicx}
\usepackage{multirow}
\usepackage{tabularx}
\usepackage{tabulary}
\usepackage{amsmath}
\usepackage{amssymb}
\usepackage{flushend}
\usepackage{color,soul}
\usepackage{textcomp}

\ifCLASSOPTIONcompsoc
	\usepackage[caption=false,font=normalsize,labelfont=sf,textfont=sf]{subfig}
\else
	\usepackage[caption=false,font=footnotesize]{subfig}
\fi

\usepackage[paperwidth=8.25in, paperheight=11in,top=2.11cm, bottom=1.2cm, left=1.4cm, right=1.4cm]{geometry}
\begin{document}

\setlength{\abovedisplayskip}{6pt}   % 0.5cm as an example
\setlength{\belowdisplayskip}{15pt}   % 0.5cm as an example

\title{A Critical Assessment of Cost-Based Nash Methods for Demand Scheduling in Smart Grids}

\author{Waleed~K.~A.~Najy,
        Jacob~W.~Crandall
	    and~H.~H.~Zeineldin,~\IEEEmembership{Member,~IEEE,}  % <-this % stops a space
%\thanks{This work was supported and funded by the Masdar Institute of Science and Technology.}% <-this % stops a space
\thanks{The authors are with the Masdar Institute of Science and Technology, Abu Dhabi, United Arab Emirates (email: wnajy@masdar.ac.ae; jcrandall@masdar.ac.ae; hzainaldin@masdar.ac.ae).}}

%\markboth{IEEE Transactions on Smart Grid}%
%{Shell \MakeLowercase{\textit{et al.}}: Bare Demo of IEEEtran.cls for Journals}

\maketitle

\begin{abstract}
%The work of Mohsenian-Rad et al. \cite{dsmnash4} has become the go-to reference on the design of an efficient demand-side management system based on game-theoretic concepts. In this article, we highlight a serious modelling flaw in the approach of \cite{dsmnash4}, as a consequence of which all the desirable theoretical results derived in the paper become invalid.

Demand-side management (DSM) is becoming an increasingly important component of the envisioned smart grid. The ability to improve the efficiency of energy use in the power system by altering demand is widely viewed as being not merely promising but in fact essential. However, while the advantages of DSM are clear, arriving at an efficient implementation has so far proven to be less straightforward. There have recently been many proposals put forth in the literature to tackle the demand scheduling aspect of DSM. One particular approach based on a game-theoretic treatment of the day-ahead load-scheduling problem has recently gained tremendous popularity in the DSM literature. In this letter, an assessment of this approach is conducted, and its main result is challenged.

%In recent years, a plethora of research on demand response in the smart grid has been produced.  The lion-share of publications on demand-response in the smart grid consists of dynamic pricing models coupled with convex optimisation solution proposals.  In this paper, we prove that such approaches are unjustified from a game-theoretic perspective.  While convex optimization techniques may have an important place in distributed power grid of the future, game theory suggests that much richer models and solutions are required.  Thus, in the second portion of this paper, we propose a new game-theoretic agenda for demand response in the smart power grid.
\end{abstract}

\begin{IEEEkeywords}
Demand-side management, smart grid, game theory, Nash equilibrium, energy pricing, multi-agent systems, load scheduling algorithms.
\end{IEEEkeywords}

\maketitle

\section{Introduction}
\label{intro}

%\subsection{Background}

\IEEEPARstart{D}{emand-side} management (DSM) is a term that refers to any attempt at increasing the efficiency of electric power use by altering the nature of the demand for power rather than its generation. DSM schemes can vary widely in complexity from the very simple, like using more efficient lighting, to the sophisticated, such as installing networked automatic controllers for home appliances. The concept of a smart grid has made proposals that fall in the latter end of this spectrum a realistic possibility, mainly because there is consensus that the future smart grid will have to rely on some form of communication infrastructure to fulfil its foreseen advantages \cite{comm}.

The goals of DSM differ depending on the point-of-view of the relevant players. Generation companies want to reduce cost and increase profit margins. Service operators, on the other hand, act as regulators on behalf of government authorities, and are concerned with safety, reliability, energy efficiency and environmental targets. Finally, customers are concerned with energy costs as well as relative sovereignty over energy use so as to maintain an acceptable standard of living. Of the above, reliability is of particular interest to all sides due to the drastic effects that loss of power can have. In the worst case, a wide-spread black-out can occur as a result of an \textit{instantaneous} shortage in supply, and this in turn would have major consequences, financial and otherwise \cite{blackout}.

The peak-to-average ratio (PAR) of the load profile represents the ratio of peak energy use at any time during the day to the average consumption. A high ratio indicates that energy use is concentrated in a particular time of day. This strains the network, since, irrespective of what the average use is, the network would have to supply at least the instantaneous demand at all times to avoid the risk of widespread failure. To this end, the most common formulations of DSM schemes have two general optimisation targets: reducing total consumption (the area under the load profile curve) as well as flattening the curve so that the peak-to-average ratio is reduced \cite{par}. These targets can, in turn, be achieved through  a variety of methods. The most frequently proposed themes are financial incentives (time-of-use rates (TOU), real-time pricing (RTP), etc.), direct control (intrusive or voluntary load control) and distributed storage \cite{dsm1}.

% ----------------------------------------
% READ THE FOLLOWING PARA AGAIN AND REVISE
% ----------------------------------------

%Shift the following to B, talk here about game-theoretic setups.

Much of the early work on investigating the performance of such methods had focused on the interaction between customers and generators. Recent articles, however, have recognised that under the DSM paradigm, it is more useful to study interactions amongst users \cite{dsmnash1}-\nocite{dsmnash2}\nocite{dsmnash3}\nocite{dsmnash4}\nocite{dsmnash5}\cite{dsmnash6}. This is because while energy efficiency metrics such as PAR are influenced by the collective actions of all customers, each household is only interested in minimising its own energy bill. Consequently, recent work has sought to align these usually conflicting goals by creating a mechanism whereby a flatter load profile is monetarily incentivised.

Of particular interest in this paper is the setup proposed in \cite{dsmnash4}, which is by far the most cited work on PAR-reducing DSM in the literature. Under this formulation, users coordinate amongst each other (rather than with the utility) to decide upon their respective day-ahead demand schedules with the aim of minimising individual energy bills under a convex pricing function. The convexity of the pricing function serves to penalise periods of high energy use with high per-unit tariffs, thus driving the system to a flatter energy profile. Due to the game-theoretic nature of the interaction, every household uses information from other households, communicated over the network, as parameters in its own cost-minimising strategy until eventually a consensus set of day-ahead schedules is reached at a Nash equilibrium.

The authors report the following theoretical results about the setup:

\begin{itemize}
\item Under the conditions of a convex pricing function and a specific per-user pricing policy, the Nash equilibrium of the system is unique, and minimises individual user energy bills.

\item Under this setup, an iterated round-robin solution method is guaranteed to converge to the unique Nash equilibrium, regardless of the initial schedules.

%\item The solution can be reached through convex optimisation, thus making the solution methodology computationally efficient. 

\end{itemize}

This work disputes the applicability of these theoretical findings by highlighting a major flaw in the assumptions used in the system. As a result, all guarantees of uniqueness, optimality and convergence are nullified.

%In this paper, an appraisal of the theoretical and practical facets of this scheme is conducted. The main limitations of the proposed scheme are proven and discussed.

%The remainder of this paper is organised as follows. In Section \ref{formulation}, we outline the mathematical formulation that implements the DSM scheme explained previously. In Section \ref{issues}, we highlight the main problem with the optimisation problems solved in \cite{dsmnash4}. In Section \ref{issues}, the convergence properties and computational efficiency of the proposed solution algorithm are investigated. Finally, conclusions are drawn in Section \ref{conclusions}.

\section{DSM Scheduling Through Cost-Based Nash Equilibria}

\label{formulation}

%\begin{figure}
%\centering
%\includegraphics[width=3.2in]{Visio-system.pdf}
%\caption{typical schematic of (a) the power system (b) the communication system}
%\label{system}
%\end{figure}

In this section, the setup and formulation under study are briefly outlined. Complete details of the setup can be found in \cite{dsmnash4}.

%One of the most cited recent work in demand-side management is \cite{dsmnash4}, where the authors propose a formulation that incentivises reducing the PAR, which is analogous to a near-flat aggregate load profile, by making it the cheaper option for users. To this end, users would attempt to schedule their electricity consumption such that each user's cost of electricity if minimised. In this section, we summarise the formulation proposed in \cite{dsmnash4} and the algorithm used to find the cost-minimising user schedules. In later sections the formulation is critiqued and the performance of the algorithm is evaluated.

%\subsection{The Demand-Scheduling Problem}

The power system under study consists of a single utility providing power to a set of users, or households, $\mathcal{K}=\{1,...,K\}$. Each household $k \in \mathcal{K}$ attempts, through an automated device, to schedule the energy demands of each its appliances $a \in \mathcal{A}_k$ in a day discretised into $H$ time periods, nominally an hour-long each. In other words, for each load of each household, the energy vector to be determined is:

\begin{align}
\boldsymbol{x_{k,a}}=[x_{k,a}^1, x_{k,a}^2, ... , x_{k,a}^h, ... x_{k,a}^H]^T
\end{align}

\noindent In addition, the user may specify that a certain appliance must be scheduled in a specific window of time during the day. Such loads are called fixed loads. Other loads may be scheduled at any time, and are called shiftable loads. 

Households are billed for electricity in the following manner: define $C: \mathbb{R}^+ \mapsto \mathbb{R}^+$ to be a strictly convex pricing function. The amount payable to the utility by each household is given by:

\begin{align}
\label{e1}
c_k = \sum_h \Bigg[\frac{\sum_a x_{k,a}^h}{\sum_k \sum_a x_{k,a}^h} \cdot C\Big(\textstyle \sum_k \sum_a x_{k,a}^h\Big)\Bigg], \forall k \in \mathcal{K}
\end{align}

\noindent The reasoning behind the pricing scheme above is the following: in each time period $h$, the total amount payable by all users collectively to the utility is $C\Big(\textstyle \sum_k \sum_a x_{k,a}^h\Big)$, which is the action of the pricing function on the total consumption of all users in time period $h$. The percentage of this amount that user $k$ must pay is given by the fraction of the total demand in that hour that he/she is responsible for. Finally, the sum of the amounts from each time period is found, thus yielding user $k$'s energy bill.

A few things are worth noting about this pricing scheme. Firstly, household $k$'s bill is a function of not only its own energy consumption but also that of all other households in the system. Thus, if we assume, accurately, that users are only concerned about minimising only their own individual energy bills (as opposed to the 'social welfare' goal of minimising the grand sum of money payable by everyone), then this immediately becomes a strategic interaction between households, and game theory comes into play. Also, because the pricing function $C(\cdot)$ is strictly convex, the effect of a linear increase in total demand in any time period by any one household, regardless of its identity, is more-than-linear in $C(\cdot)$, and thus the effect is an increase in the energy bill of any one user - the linear increase in the denominator of (\ref{e1}) - is offset by a more-than-linear increase in $C(\cdot)$). In a sense, the pricing scheme above implements a form of collective punishment that aims at pushing the system towards flatter profiles of total demand, which would minimise $C(\cdot)$.

Each user would then aim to minimise $c_k$ in a distributed, game-theoretic manner. A centralised optimisation problem would be computationally ideal, but would rob users of sovereignty over their schedules, and would also violate the privacy of users. Thus, the problem must be solved in a distributed fashion, with each household deciding upon its own schedule.

Under such a distributed setting, a convergence point of the system is the Nash equilibrium, i.e., the point at which the following holds:

\begin{align}
 c_k(\boldsymbol{x^*_k},\boldsymbol{x^*_{-k}}) \leq  c_k(\boldsymbol{x_k},\boldsymbol{x^*_{-k}}), \forall k
\end{align}

It is hypothesised in \cite{dsmnash4} that under the setup described above, the Nash equilibrium is unique, and also corresponds to the system state that globally minimises cost for every user. In addition, the authors propose an algorithm that converges to this Nash equilibrium. All users randomly initialise their load schedules. Then, on user $k$'s turn, all other users broadcast their current schedules to user $k$ (in an aggregated fashion to avoid privacy issues), who uses these schedules to generate an updated schedule by solving the following local cost-minimisation problem using convex optimisation:

\begin{align}
\nonumber \underset{\boldsymbol{x_k} \in \mathbb{R}_+^{H \times |\mathcal{A}_k|}}{\text{minimise}} &\hspace{12pt} \sum_h \Bigg[\frac{\sum_a x_{k,a}^h}{\sum_k \sum_a x_{k,a}^h} \cdot C\Big(\textstyle \sum_k \sum_a x_{k,a}^h\Big)\Bigg] \\
\nonumber \label{e3} \text{subject to} &  \hspace{12pt}   \sum_h x_{k,a}^h = E_{k,a}, \forall a \in \mathcal{A}_k \\
 & \hspace{12pt} x_{k,a}^h = 0, \forall h \notin \{\alpha_{k,a},\cdots,\beta_{k,a}\}
\end{align}

\noindent where $E_{k,a}$ is the predetermined energy demand for user $k$'s appliance $a$, and $\alpha_{k,a}$ and $\beta_{k,a}$ are the starting and ending times of appliance operation. We note that the above optimisation problem is solved over user $k$'s demand only, with the demands of all other users being parameters to user $k$'s local problem. When all users do this, one after another, passing forward their updated schedules in each iteration, users will converge to a set of schedules where no more updates occur, and it is suggested that this set of schedules represents the unique Nash equilibrium of the scheduling game. We note that this method of finding the Nash equilibrium is in fact the Cournot adjustment model \cite{fudenberg}, where firms strategise in a sequential manner, and the strategy chosen is the best response to the action of opposing firm.

\section{Issues with the Current Formulation}
\label{issues}

\subsection{Unrealistic Modelling Assumptions}

The fundamental issue with the approach of \cite{dsmnash4} is that for optimisation problem (\ref{e3}) is solved in the \textit{real space} $\mathbb{R}^{H \times |\mathcal{A}_k|}$. This represents a critical modelling error, since a typical household appliances has a fixed power rating, meaning that it will consume a fixed, discrete amount of energy in each time period when it is on. Because the total energy consumption of an appliance is assumed to be known \textit{a-priori}, the total length of time the appliance must be on is also known. Thus the scheduling problem reduces to one where the only decision variable per appliance that needs to be solved for is the time at which the appliance must be turned on. This is, of course, a discrete problem, and this has important implications regarding the hardness of solving the optimisation problem at hand. Solving in the real space, however, implies that we have complete control over the power rating of each appliance, which is highly unrealistic.

This fact alone is enough to nullify all the theoretical results derived in \cite{dsmnash4}.

\subsection{Non-Uniqueness of the Nash Equilibrium}
\label{non-uniqueness}

%\textit{Proposition 1: The above formulation does not guarantee a unique Nash equilibrium for the demand-scheduling game.}

The non-uniqueness of the value of the Nash equilibrium can be shown through a counter-example. Assume a scenario, which we will refer to as Scenario I, in which two users, $U_1$ and $U_2$ attempt to schedule loads in a day with a time granularity of 3, i.e., with $t \in \{1,2,3\}$. User $U_1$'s demands consist of one 1kWh per time period (kWh/tp) fixed load that must be scheduled during $t=1$ with a duration of one time period, and one 2kWh/tp shiftable load whose duration is one time slot, and may be scheduled in any of the three time slots. User $U_2$ has one 5kWh/tp fixed load with a duration of one time period in $t=2$ and one 2.5kWh/tp shiftable load with a duration of one time period. Assume also that the convex cost function used is $C(\cdot)=(\cdot)^2$ cents.

\begin{table}
\centering
\normalsize
%\footnotesize
%\scriptsize
\caption{The payoff matrix for Scenario I}
\begin{tabular}{c|c|c|c}
	\hline
	            &               $t_{U_2}=1$               &          $t_{U_2}=2$           &               $t_{U_2}=3$               \\ \hline
	$t_{U_1}=1$ &     15.79\textcent, 39.46\textcent      & 18.64\textcent, 46.61\textcent & \textbf{11.50\textcent, 28.75\textcent} \\ \hline
	$t_{U_1}=2$ &     17.50\textcent, 43.75\textcent      & 26.07\textcent, 65.18\textcent &     16.07\textcent, 40.18\textcent      \\ \hline
	$t_{U_1}=3$ & \textbf{11.79\textcent, 29.46\textcent} & 17.50\textcent, 43.75\textcent &     13.21\textcent, 33.04\textcent      \\ \hline
\end{tabular}

\label{games}
\end{table}

The two-player, three-action game for this scenario, obtained by applying Eq. (\ref{e1}), is shown in Table \ref{games}. Each cell of the matrix shows the cost for each user for the indicated outcome. For example, when player 1 selects time-period $t=2$ and player 2 select time-period $t=1$, player 1 incurs a cost of 17.50\textcent { and} player 2 incurs a cost of 43.75\textcent. In each of the joint outcomes in bold, no user has an incentive to switch the time of usage of his shiftable load given that the other player will maintain his current schedule because he would have to pay a larger amount. Thus, there are two pure-strategy Nash equilibria in this scenario.  Therefore, it is clear that this formulation does not guarantee a unique Nash equilibrium.

%It is essential to note that the system is discrete not only in time but also in energy consumption (and consequently in cost), since appliances have fixed power ratings and will therefore consume a fixed amount of energy in each time period. As such, the only problem variable is the starting time period of each load. In addition, optimising in continuous time and energy domains also yields these same two Nash equilibria. This is because scheduling any part of the shiftable load in time period $t=2$ is never the cheaper option due to the presence of the large 5kWh fixed load in that slot.

%--------------------------------@@@@@@@@@@@@@@@@@@@@@@@@@@@@@

\subsection{Convergence to Non-Pareto-Optimal Nash Equilibria}
%\textit{Proposition 2: The Cournot adjustment algorithm is not guaranteed to converge to the Pareto-optimal Nash equilibrium.}

In Table \ref{games}, the Pareto-optimal Nash equilibrium is ($t_{U_1},t_{U_2})=(1,3)$ since this achieves the minimum-cost state, (11.50\textcent, 28.75\textcent), for both users. If, however, the users randomly initialise their schedules to the joint action ($t_{U_1},t_{U_2})=(3,1)$, it would not be possible to escape the sub-optimal solution (11.79\textcent, 29.46\textcent) with the Cournot adjustment model since each user would incur a higher cost for a unilateral change of action. Thus, the two users converge to the sub-optimal Nash equilibrium. This in turn leads to sub-optimal reductions in PAR and cost.

A similar analysis was performed on a six-player game in Scenario II, with time granularity of six slots, a total of 2kWh/tp of fixed loads over three time-slots starting at $t=4$, and one shiftable load per user of sizes $[1.5,2,0.67,2,5,0.33]$ kWh/tp of durations $[1,2,3,1,2,3]$ time periods respectively. Thus, six users attempt to schedule their energy demands in a day-ahead fashion, with each agent aiming to minimise its own energy bill. Each user is made to have only one shiftable load to facilitate clarity and ease of exposition of the results without loss of generality. The same iterative algorithm implemented in Scenario I is again used here by each user to find the starting times of shiftable loads.

\begin{table}[t]
\centering
\normalsize
\caption{Distribution of convergence to various Nash equilibria}
\begin{tabular}{c|c|c|c} \hline
\textbf{No. of NEs } 		& $\sum_h C(X^h)$ & \textbf{ \% Convergence to NE} & \textbf{PAR}\\ \hline
4	&	157.58\textcent		&	5.70\%		&		1.11 \\ \hline
7	&	157.81\textcent		&	11.02\%		&		1.11 \\ \hline
8	&	158.25\textcent		&	14.05\%		&		1.18 \\ \hline
2	&	158.69\textcent		&	11.08\%		&		1.18 \\ \hline
16	&	158.92\textcent		&	41.68\%		&		1.18 \\ \hline
12	&	162.47\textcent		&	16.47\%		&		1.25 \\ \hline
1	&	163.58\textcent		&	0\%			&		1.38 \\ \hline 
\end{tabular}
\label{sc2results}
\end{table}

\begin{figure}[t]
\centering
\includegraphics[width=5.5in]{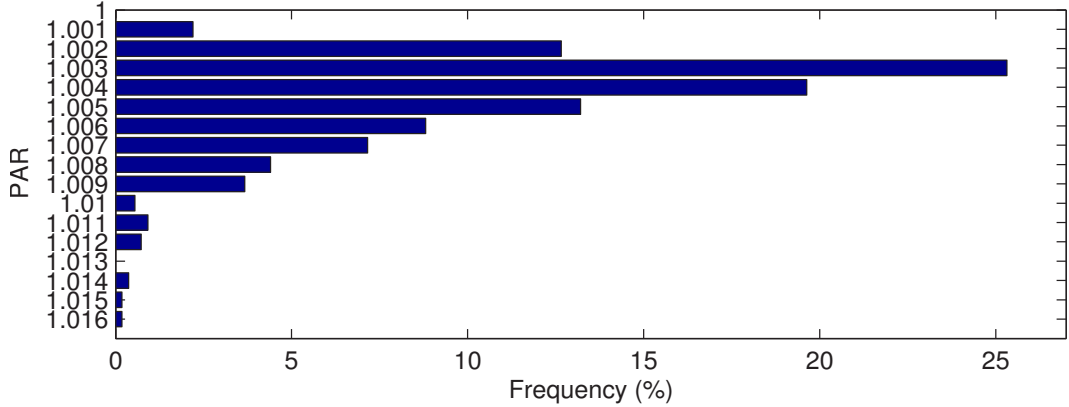}
\caption{Distribution of PARs obtained in Scenario III}
\label{sc3}
\end{figure}

If loads are allowed to overflow, or loop, to the next day then there are $6^6$=46,656 possible joint outcomes. An exhaustive search reveals that 50 of these outcomes are Nash equilibria, thus illustrating again the non-uniqueness of the Nash equilibrium in the system. Empirical analysis was performed with 100,000 runs of simulation, each with random initialisation of loads at the start of the algorithm. The results are shown in Table \ref{sc2results}. The first column shows the joint combinations of load starting times that correspond to Nash equilibria. The second column shows the value of $\sum_h C(X^h)$ for each of these states. Clearly, some Nash equilibria share the same total cost. It is worth reiterating that because all derived games are purely cooperative, it suffices to state the total cost, $\sum_h C(X^h)$, of the state since this quantity is directly proportional to the preference of all users for the state. It can be seen that using the iterated Cournot dynamic, convergence to the Pareto-optimal Nash equilibrium occurs only 5.7\% of the time Also, the optimal PAR is achieved only 16.7\% of the time.

A final case-study is considered (Scenario III) that is very similar to the one used in \cite{dsmnash4}. The system consists of 10 customers, each with 20 flexible loads, optimising cost in a day of hour-long time periods ($H$=24). The system was designed to have several solutions that yield a perfect PAR of unity. Due to the huge size of the problem (there are 200$^{24}$ possible outcomes), it is not possible to identify the number of Nash equilibria in this system. However, simulation gives an empirical distribution of Nash equilibrium PARs, shown in Fig. \ref{sc3}. The results confirm the statements made previously regarding the non-uniqueness and non-optimality of the achieved, but also demonstrate that the algorithm performs quite well in this scenario by finding a Nash equilibrium that achieves a PAR at most 2\% away from the ideal value almost all the time. Thus, while the algorithm is not guaranteed to converge to a Pareto-optimal NE, it sometimes does.  This empirical analysis suggests the encouraging result that the algorithm is more likely to find near-optimal solutions in larger systems than in smaller systems.

\section{Dynamics of Repeated Play with Sovereign Actions: Further Issues}
\label{repeatedplay}

The distributed nature of the system offers interesting advantages to players. Because decision-making is decentralised, self-interested users are free to strategise in an effort to maximise their own utilities. The formulation described in Section \ref{formulation}, however, is very restrictive in that regard. It can be noted that while users attempt, and succeed to some degree, in reducing their costs, they do not have a real choice to make in the system. It goes without saying that users will always choose a lower cost over a higher one. In other words, it is not clear what the added value of the decentralised computing of equilibria is when compared to having all users send their initial schedules to the utility and have it perform the computation. In addition, the larger flow of information in the network adds to the computational burden of the system, which becomes more of a problem as the system scales to more users \cite{dsmnash5}.

If, on the other hand, users were to have some degree of freedom to strategise, users may have the ability to achieve larger expected savings over time. However, if a system employing the formulation in Section \ref{formulation} enables users to override the outcome of the algorithm, it quickly becomes apparent that such a scheme is not lying-proof. For example, users who decide to report certain shiftable loads as fixed ones to ensure that they attain their preferred time periods could bully other cost-minimising users away from these time periods. While it is true that other users can do the same in response, differences in the non-monetary valuations for time slots by users would not be captured by the cost function. This would lead to disparities in the resulting game, and an incentive to lie can arise. In this section, we discuss further pitfalls in the system in question that naturally arise as a result of repeated interaction over many days.

\subsection{Game-Changing Valuations of Time Periods}

\begin{figure*}[t!]
\centering

\begin{tabular}{c}

\begin{tabular}{c||c|c}
\multicolumn{3}{c}{(a) Costs} \\
\hline
		&	$t_{U_2}=1$ & $t_{U_2}=2$ \\ \hline \hline
$t_{U_1}=1$ & -2, -2 & {\bf -1, -1 }\\ \hline
$t_{U_1}=2$ & {\bf -1, -1} & -2, -2 \\ \hline
\end{tabular} 

\begin{tabular}{c}
\Large{+}
\end{tabular}

\begin{tabular}{c||c|c}
\multicolumn{3}{c}{(b) Valuations} \\
\hline
		&	$t_{U_2}=1$ & $t_{U_2}=2$ \\ \hline \hline
$t_{U_1}=1$ & 3,3 & 3,1 \\ \hline
$t_{U_1}=2$ & 1,3 & 1,1 \\ \hline
\end{tabular} 

\begin{tabular}{c}
\Large{=}
\end{tabular}

\begin{tabular}{c||c|c}
\multicolumn{3}{c}{(c) Effective Game} \\
\hline
		&	$t_{U_2}=1$ & $t_{U_2}=2$ \\ \hline \hline
$t_{U_1}=1$ & {\bf 1,1} & 2,0\\ \hline
$t_{U_1}=2$ & 0,2 & -1,-1 \\ \hline
\end{tabular} \\

\end{tabular}
\caption{Effect of slot preferences on the scheduling game}
\label{games2}
\end{figure*}

While it is insightful to formulate interactions between users on the basis of cost, it is also an incomplete approach. User valuations of running certain loads in certain time-periods can have a significant effect on the characteristics of the load-scheduling game. Certain users may, for example, value the convenience of being able to run the washing machine during the day enough to be willing to pay the extra cost. This effect would be amplified even further if there are substantial differences in affluence among the users involved. In such a case, it is possible that the Nash equilibrium of the game changes such that it lies in a state that exasperates the PAR rather than reduces it.

To illustrate this, consider the game in Fig. \ref{games2} (a). This is very similar to the game in Fig. \ref{games} (a); there are two Nash equilibria and they occur when users schedule their shiftable loads in different time slots, hence reducing the PAR. If, however, each user values time-period $t=1$ as 3 payoff units and $t=2$ as 1 payoff unit, the matrix in Fig. \ref{games2} (b) is added to the original game, yielding the new game in Fig. \ref{games2} (c). In this game, however, the Nash equilibrium occurs when both users schedule their shiftable loads in slot $t=1$, which results in a larger PAR. This simple example illustrates how preferences over time periods for load consumption can lead to a situation in which the resulting outcome is the opposite of the intended purpose.

\subsection{Strategising in Repeated Play}

\begin{figure}[t]
\centering
\includegraphics[width=3.5in]{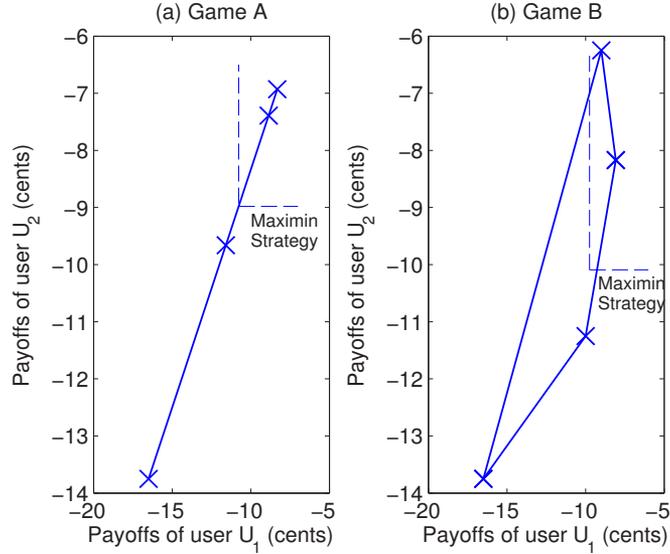}
\caption{Convex hulls of payoffs of Game A and Game B in Fig. \ref{games}.}
\label{convexhulls}
\end{figure}

The dynamics of repeated play in the demand scheduling game have so far been ignored in the literature, mainly due to conclusions about the presence of a unique one-shot Nash equilibrium in the game. While strategic play may not offer any significant added value to any player in purely cooperative games, repeated play becomes an interesting problem in games of conflicting interest \cite{noncoop}, such as those derived from pricing scheme B, where the joint outcome that minimises the user's costs are not the same for all players. This is true even in the relatively simple case of coordination games \cite{coord} - ones in which there is a single joint outcome that is optimal for every user, but where preferences of other states are not necessarily the same. The difference is visually clear in Fig. \ref{convexhulls}. The most obvious source of complication is the possibility of ``running around in circles" when using myopic strategies that implement the immediate best response. In other words, users can keep jumping from one sub-optimal outcome to another, thus reducing the average outcome of the repeated game in the limit.

Another problem with myopic best-response strategies is intimately related to concepts of social welfare and fairness, and stems from the well-known game theory concept known as the folk theorem \cite{folk}. The folk theorem states that in repeated play, every point that lies in the convex hull of joint outcomes of the game and that Pareto-dominates the maximin strategy \cite{gametheory} can be sustained as a Nash equilibrium of the repeated game. This implies an infinite number of Nash equilibria in repeated play. This also allows room for bullying strategies, where users can prosper by lying about their preferences.

Strategic play comprises a whole body of work in the artificial intelligence literature, and a thorough discussion of these concepts is therefore out of the scope of this work. It suffices here to point to the fact that this area of research offers a rich source of potential solutions when more interesting games are involved, and that the smart grid community would greatly benefit from leveraging this body of research.

\section{Discussion of Possible Alternative Frameworks for DSM}
\label{alternative}

The algorithm proposed in \cite{dsmnash4} and many others is a very useful proposal for implementing demand-side management. It incentivises PAR reduction by linking it to a cut in expense for all users in the system simultaneously. Also, empirical tests of the algorithm show that it converges very quickly even for games with a large number of agents. However, it was proven in the previous sections that the theoretical results on which the performance of the solution algorithm hinges are in fact inaccurate. The question at this point is whether it is possible to devise an alternative algorithm under the same DSM paradigm that performs better.

Firstly, the intrinsic limitations of the algorithm that were discussed in Section \ref{issues} must be eliminated. This appears to be a difficult goal to achieve. To begin with, there are simply no algorithms that are known to efficiently guarantee good payoffs in games with a large number of agents. The space of joint actions increases exponentially and the curse of dimensionality becomes overwhelmingly difficult to overcome. To see this, consider a system where $M$ appliances must be scheduled, and where the number of time slots per day is $N$. The number of possible joint states is $N^M$. Of these, only a handful of joint states would represent Pareto-optimal Nash equilibria. Therefore, the probability of reaching such a state using a Cournot dynamic is of the order of $N^{-M}$, which clearly tends to zero very quickly, even for relatively small values of $M$. Another problem is that when the scheduling game grows larger, the number of (suboptimal) Nash equilibria also increases quickly. Intuitively, this is due to the fact that the same payoffs can be achieved by simple rotations of solution schedules that yield similar costs to all users.

Nevertheless, some pragmatic approaches appear to be worth testing. For example, the utility can mandate that loads that are larger than a particular threshold must be scheduled by the utility. These are the loads that affect the load profile most considerably, and through such regulation, the utility can first solve the much smaller, yet more significant, problem of scheduling large loads. The result of this optimisation can then be passed on to the users, who would use this result as an input parameter to the load-scheduling game. The fact that the most significantly-sized loads are already scheduled would make the space of Nash equilibria much smaller. Equivalently, it becomes possible for users to prune the search-space, or at least prioritise over it, by considering less congested slots first. Conceptually, the smaller loads would fine-tune the load profile by plugging gaps in it with a reduced risk of overshooting the target of a flat profile (minimum PAR) due to the smaller energy requirements of the remaining loads. The reduced dimensionality of the resulting scheduling game also aids the efficiency and convergence of any implemented scheduling algorithm.

From a game-theoretic perspective, a reinforcement learning approach also seems promising. If we assume that the energy demands of users are more-or-less constant in terms of type and size over an extended period of time, consumers can track the payoffs of their actions and adjust their schedules to ones that have historically proven to be profitable. More specifically, the well-known Q-learning algorithm \cite{qlearning} can be used to choose a schedule that minimises the expected cost based on historical data of the cost incurred by the user when such a schedule was chosen in the past. While Q-learning is known to be slow to converge, especially in games with many agents, it could offer an improvement on the approach proposed above.

The issues raised in Section \ref{repeatedplay} are also difficult to address. There have been attempts at quantifying user valuations in the context of the DSM demand scheduling problem using quasi-linear cost functions \cite{mechanism}. Such functions attempt to explicitly capture all aspects of user valuation functions that do not include the financial cost of any given choice. However, it remains unclear how accurate any assumed valuation function would be in a practical setting. 

%\section{Simulations and Results}

%\begin{figure}[t]
%\centering
%\includegraphics[width=3.5in]{sim1.pdf}
%\caption{Simulation results from a two-player 50-action game showing the frequency of convergence to every state state.}
%\label{sim1}
%\end{figure}
%
%Fig. \ref{sim1} shows simulation results obtained from a two-player 120-action game generated from a simple but realistic Scenario II described in Appendix \ref{app2} according to pricing scheme (a), where each user implements the iterative algorithm described previously to arrive at a schedule. The game was simulated 10,000 times. It is clear that there are several Nash equilibria to which the iterative algorithm can converge. Moreover, convergence to the Pareto-dominant Nash equilibrium was recorded only 28\% of the time on average.

\section{Conclusions}
\label{conclusions}

In this paper, a common DSM formulation based on finding the Nash equilibria of load-scheduling games derived from the energy costs of users was described and critically assessed. After outlining the mathematics of the scheme, several issues pertaining to PAR reduction and computational performance were discussed. In particular, it was shown that due to the non-uniqueness of the Nash equilibria of load-scheduling games, the proposed algorithm was not guaranteed to converge to the PAR-minimising solution. It was also shown that convex optimisation was not the most suitable tool for attempting to find such an equilibrium point due to the nature of the solution space. In addition, it was explained that the performance of the setup is further worsened in situations where users can override the schedules calculated by the scheduling algorithm, and when there is no clear joint outcome that is beneficial to all users simultaneously. Furthermore, it was shown that when the games are influenced by differential valuations of time slots by users, the algorithm may in fact converge to a point that aggravates the PAR rather than improve it.

\appendices

\bibliographystyle{IEEEtran}
\bibliography{IEEEabrv,dsmbib}

\atColsBreak{\vskip-5pt}

\end{document}